\documentclass[aps,superscriptaddress,twocolumn,showpacs,preprintnumbers,amsmath,amssymb,showkeys]{revtex4}
\usepackage[cp1251]{inputenc}
\usepackage[english]{babel}
\usepackage{amssymb,amsmath}
\usepackage{amstext} 
\usepackage[dvips]{hyperref}
\usepackage[mathscr]{eucal}
\usepackage{longtable}
\usepackage{graphicx}
\begin{document}
\title{Quantum Fermi Liquid Decription of (Quasi)-One-Dimensional Electronic  Systems}
\author{Alla Dovlatova (a), Dmitri Yerchuck (b), Felix Borovik (b)\\
\textit{(a) - M.V.Lomonosov Moscow State University, Moscow, 119899, \\ (b) Heat-Mass Transfer Institute of National Academy of Sciences of RB, Brovka Str.15, Minsk, 220072, Corresponding author is Dmitri Yerchuck, E-mail: dpy@tut.by}}
\date{\today}%
\begin{abstract} The concept of quantum Fermi liquid for description of (quasi)-1D electronic systems is recovered. The model of (quasi)-1D quantum Fermi liquid is developed on the example of \textit{trans}-polyacetylene and it is   the generalization of well-known  model  of organic (quasi)-1D conductors, elaborated by Su, Schrieffer and Heeger (SSH-model). It is shown, that spin-charge separation effect can be realized in (quasi)-1D quantum Fermi liquids.  It has topological soliton  origin in distinction from   spinon-holon spin-charge separation effect in Tomonaga-Luttinger liquids and electronic systems like them. The model  allows to extend the limits of the applicability of SSH-model to  the electron-electron correlated  (quasi)-1D-systems without restriction on electron-electron interaction force. The (quasi)-1D-systems with strong electron-phonon interaction and/or strong electron-photon interaction can be also described within the framework of given 1D quantum Fermi liquid model. Practical significance of the model proposed consists in the clarification of the nature of charge and spin carriers and in the clarification of  the origin of mechanisms of quasiparticles' interaction in  the  (quasi)-1D-systems, which are the objects of nanoelectronics, spintronics and the other modern nanotechnology branches.

\end{abstract}  

\maketitle                         
\section{Introduction} There seems to be very essential  for the tasks of nanoelectronics, spintronics  and for the other branches of nanotechnology the knowledge  of the nature  of charge and spin carriers in the nano-devices. Especially seems to be  significant the knowledge of mechanisms  of carrier transport and interactions of charge and spin carriers  both between themselves and with phonons and  photons. In particular, great hopes are pinned in modern nanotechnology on carbon nanotubes (NTs), that is, in using of carbon NTs for production of the main  devices for nanoelectronics and related nanotechnology  branches.

  There is existing in the theory of (quasi)-1D electronic systems, in particular, in the theory of conducting NTs, the following concept, which was starting with the work of Tomonaga in
1950 \cite{Tomonaga} and  with the work of Luttinger in 1963 \cite{Luttinger_J},  when it has become clear that the electron-electron
interaction destroys the sharp Fermi surface and leads to a breakdown of the
Landau  Fermi liquid (LFL) theory.  The resulting non-LFL state is commonly called Luttinger
liquid (LL), or  Tomonaga-Luttinger liquid (TLL).   It was suggested, that  TLL describes the universal low-energy properties of
 1D conductors. TLL behaviour is characterized  theoretically by
pronounced power-law suppression of the transport current and the density of states,
and by effect of spin-charge separation. The idea  of spin-charge separation was originally
introduced by Anderson in 1987 \cite{Anderson}, \cite{Anderson_P_W}, \cite{Anderson_Ren} for doped Mott-Hubbard insulator in the metallic regime. Similar spin-charge separation effect was mathematically realized in
the so-called slave-particle representation  \cite{Zou_Anderson} of the $t-J$ model. The nature of the spin and charge carriers  in the TLL concept  is characterised by the folllowing. They are in Anderson approach chargeless spin 1/2 quasiparticles - spinons and spinless quasiparticles with the charge $\pm{e}$   - holons. 
  So Anderson spin-charge separation effect  may be mathematically realized in
the so-called slave-particle representation  \cite{Zou_Anderson} of the $t-J$ model
$e_{i\sigma} = h^+_i f_{i\sigma}$, 
where $h^+_i$, $f_{i\sigma}$ are holon  and spinon fields. The occupancy constraint, reflecting the Hubbard gap in its extreme limit, is handled by an equality
 $h^+_i h_i + \sum_{(\sigma)}{f^+_{i\sigma}f_{i\sigma}} = 1$, 
 which commutes with the Hamiltonian. It is seen the close relation of the spin-charge
separation and the constraint condition through the counting of the quantum numbers. But the spin-charge separation  acquires a new meaning here. If those holon 
 and spinon  fields indeed describe elementary excitations,
the hole (electron) is no longer a stable object and must decay into a holon-spinon pair once being injected into the
system. This instability of a hole (electron), being to be free quasiparticles  in solid state physics theory (within the frames of effective mass method) is referred in the literature, see, for instance, \cite{Weng} to be the deconfinement, in
order to distinguish it from the narrow meaning of the Anderson mechanism of spin-charge separation about elementary excitations.

 TLL description is argued to be universal. The universality of TLL description means in its turn, that the physical properties do
not depend on details of the model, the interaction potential, and so on, but instead they
are only characterized by a few parameters - critical exponents. Moreover, the TLL concept is
believed to be true for arbitrary statistical properties of the particles, that is,  both for
fermions and bosons. It provided a paradigm for non-Fermi liquid physics.
 
Let us remark, that it  is argued in many works,  that 
 the single-wall carbon nanotubes (SWCNTs), considered to be 1D objects (it is not always correct, especially for standard NTs with diameter in several nanometers)  can
be described the only within the frames of  TLL concept. Moreover, SWCNTs
are considered  to be the best model system of the TLL state demonstration. Given viewpoint does not have, however, reliable experimental   grounds.  Really,
 power-law behaviour was observed experimentally by measuring
the tunneling conductance of SWNTs in dependence on  temperature and voltage.  Electron force microscopic measurements showed also the
ballistic nature of transport in conducting SWNTs.  At the same time, the most prominent feature of the model - spin-charge separation by spinon-holon mechanosm - has not been observed
so far. Moreover, power-law behaviour of conductance of SWNTs in dependence on  temperature and voltage and ballistic nature of transport are rather universal physical properties. They can be obtained within the frameworks of the other theories. Consequently, we have to conclude, that the existing viewpoint seems to be insufficiently experimentally confirmed.

\section{spin-charge separation - brief review}

When concerning the other 1D systems, spinon-holon mechanosm of spin-charge separation seems to be also not proved experimentally. Only a few experiments have attempted to
detect the spin-charge separation in (quasi)-1D systems directly, \cite{Kim}, \cite{Segovia},\cite{Claessen}, \cite{JKim},\cite{Dudy}, angle resolved photoemission spectroscopy (ARPES) was used with given aim. Let us analyse the explanation of the results in the  work \cite{Dudy}, where the attempt is undertaken to prove experimentally the existence of spinon-holon spin-charge separation quite recently. Angle resolved photoemission spectroscopy  studies in quasi-1D compound   $Li_{0.9}Mo_6O_{17}$ have shown, that observed  lineshapes are asymmetric. Authors ascribe the asymmetry observed to manifestation of spin-charge separation by holon-spinon mechanism. However, they have observed the only one-peaked spectrum. At the same time two-line spectrum is predicted by realisation of given mechanism \cite{Martino}. It has been done in \cite{Martino} for radiospectroscopy range, however, qualitatively, by ARPES measurements the same picture has to be observed. Authors of \cite{Dudy} believe, that  there is the only  dispersing holon peak and
spinon edge instead of peak. In given suggestion, the observed lineshape can be partly fitted by the ratio of holon $v_c$ to spinon $v_s$ velocities, equal to 2. Let us remark, that the propagation of the electrically charged quasiparticles with  the velocity, strongly exceeding the velocity of electrically neutal spin 1/2 quasiparticles contradicts the results, obtained for many quasi-1D systems, for instance, for \textit{trans}-polyacetylene (t-PA) \cite{SSH}, \cite{SSH_PRB}, \cite{Heeger}, carbynoids \cite{Ertchak_J_Physics_Condensed_Matter}, quasi-1D NTs \cite{Ertchak_JAS},\cite{Yerchuck_Dovlatova}. It has been shown in above cited works \cite{SSH}, \cite{SSH_PRB}, \cite{Heeger}, \cite{Ertchak_J_Physics_Condensed_Matter}, \cite{Ertchak_JAS},\cite{Yerchuck_Dovlatova}, in which the spin-charge separation effect is surely experimentally established, that electrically neutal spin 1/2 quasiparticles
are very mobile, whereas, charged spinless quasiparticles can at all be pinned \cite{Heeger}. Admittedly, the spin-charge separation effect has quite another origin in given case (see the next Section). Much more simple explanation for ARPES lineshape observed in \cite{Dudy} can be proposed. It can be, for instance, the realisation of opical analogue of Dyson effect \cite{Dyson}. The second very possible explanation of the line asymmetry observed consists in  the appearance of distribution of hot photoelectrons in kinetic energy by their moving to the surface. It is seen from Figure 5 in \cite{Dudy}, that it  corresponds well to the presence of energy distribution of the particles in Fermi gas, (or quasiparticles in Fermi liquid). Let us remark, that quite similar asymmetric lineshapes of the lines were observed by optical absorption on excitons, condensed in Fermi liquid in a number of dielectrics and semiconductors and given result was explained theoretically just by  distribution of moving excitons in kinetic energy  in very good agreement with experimental lineshapes.

Therefore, the connection of the asymmetry of lineshapes in ARPES measurements with spinon-holon charge separation seems to be not proved in the works above cited.

It has to be also remarked, that both the models TLL and LFL  are the models of ideal quantum liquids, moreover, they are  oversimplified,  since they do not take into account the nonlinearity of the fermion spectrum on the one hand and it is especially critical, that they do not take into account relaxation mechanisms, including mechanism, realised by means of electron-phonon interactions, on the other hand. In fact, both the models describe not strongly adequately the real physical  processes (like to image in distorting mirror). Actually,  the changes in a charge state of arbitrary atom in 1D chain in the result of electron-electron interaction are always accompanied by the changes in phonon subsystem (and vice versa). It is consequence of generic coupling between operators of creation and annihilation in electron subsystem and in phonon field (see for details the next Section). Consequently, the models, which do not take into account the electron-phonon interaction, seem to be strongly  oversimplified and very restricted models in its applications. They can never explain the full set of experimental results. It is useful to remark, that the development of 1D TLL model by means of including of electron-phonon interaction has been done in  \cite{Ning} through the calculation
of one-electron spectral functions in terms of the  cluster perturbation theory
together with an optimized phonon approach.  It was found, that the retardation effect, which is the consequence  of the
finiteness of phonon frequency, suppresses the spin-charge separation by spinon-holon mechanism and eventually makes it invisible
in the spectral function. 
It is strong additional indication, that the fitting of spectral function on the base of TLL theory by authors of \cite{Kim}, \cite{Segovia}, \cite{Claessen}, \cite{JKim}, \cite{Dudy} can be physically incorrect. 

Quantum critical (QC) scaling in the lineshapes has been also studied in \cite{Dudy}. The authors themselves have been found the departures of the 
lineshapes and the scaling from TLL-theory and substantial
differences relative to expectations from the one-band TLL-model, specifically  in the
exponent of the temperature prefactor and in the lack of the full  sharpening predicted by TLL-theory
for decreasing of observation tempeature. The authors have phenomenologically included momentum
broadening of the TLL spectral function, origin of which remained to be unknown.
The discrepancies became smaller, however, they are remained.
 
So we see, that  reliable experimental confirmation of spinon-holon spin-charge separation mechanism in quasi-1D systems by means of ARPES is in fact absent. Given viewpoint is in agreement with the opinion of the authors of \cite{Jompol}. They remark, that  in these experiments
TLL is "both probe and subject, so an independent study - in a different geometry - of
the excitation spectrum is vital to be sure of the interpretation". Given
remark of the authors of \cite{Jompol} refer also to the works 
\cite{Auslaender}, \cite{Tserkovnyak}, \cite{AuslaenderO}, in which the authors discuss the evidence for spinon-holon spin-charge separation in tunneling between
two parallel quantum wires at a cleaved edge of a double–quantum-well heterostructure. For instance, in \cite{Tserkovnyak}, 
 two approaches have been used - one based on mapping out the elementary-excitation dispersions by
measuring the conductance $G$ in dependence on the magnetic field $B$ applied perpendicular
to the plane connecting the wires and the voltage bias $V$, and the other focusing on the
conductance oscillation pattern, in the $(V,B)$ plane, arising to be the result of the finite length of the
tunnel junction.

We have to remark, that   the authors of \cite{Auslaender}, \cite{Tserkovnyak}, \cite{AuslaenderO}  and the authors of \cite{Jompol} too were dealing with element of 2D system, since 1D-1D tunnelling process determines the second direction, being to be transversal to the direction, along which two parallel quantum wires were placed. It means, that TLL model has to be generalized for 2D case before its numerical application for the explanation of the results obtained. It is in principle correct task. Really,
Haldane \cite{Haldane} has presented a generalization of the "bosonization" description, which is key
moment in TLL model, to be a general treatment of Fermi
surface dynamics in any dimension. Generally, from mathematical viewpoint, Luttinger liquid behaviour will be observed independently on the dimensionality for the systems, for which the energy at
Fermi surface is not extremal and, consequently, the linear term has to be
preserved in its Taylor expansion about the Fermi surface points. It is understandable, that, if to restrict the consideration of the task with accuracy 
to linear term in Taylor expansion, than we obtain Luttinger liquid behaviour for massless quasiparticles, since
effective mass is appeared in quadratic term by Taylor
expansion.  It really takes
place, for instance, for graphene, when to consider the only
 nearest-neighbour hopping interaction between the
atomen. 

Therefore, the interpretation of the results, obtained by tunneling between
two parallel quantum wires or on related systems has to take into consideration given remark. A priori, we can suggest, taking into account the description of physical properties of  graphene, that mathematical treatment will be different  in comparison with the treatment, proposed in \cite{Auslaender}, \cite{Tserkovnyak}, \cite{AuslaenderO}, \cite{Jompol}. [Nevetheless, the experimental results, presented in given works, seem to be very interesting].
 
Let us give some details, concerning the description of the physical properties of graphene. 
In pristine graphene, the Fermi level lies just at the touching (crossing) point (the Dirac or charge neutrality
point) of $\pi^*$ 
and $\pi$  bands and graphene has a character of zero-band-gap semiconductor (semimetal). Band structure on some distance from Fermi level in a standard
tight-binding approach and by considering  the only the nearest-neighbour hopping consist of six symmetric Dirac cones with vertices, which produce regular hexagon \cite{Castro}.
Close to a given crossing (touching) point, the electronic bands are nearly linear and practically rotationally symmetric. In
other words, the carrier dispersion relations take a simple form
\begin{equation}
\label{Eq16q} 
E_\pi^* = - E_\pi \approx v_F \hbar |\vec{k}|,
\end{equation}
where the momentum  $\vec{k}$ is measured with respect to $K (K')$ point. The parameter $v_F$,
having dimension of a velocity, is directly related to the coupling strength (hopping
integral) between the nearest carbon atoms: $v_F = \sqrt{3} a_0 \gamma_0/(2 \hbar)$, where $a_0$ is lattice constant, $\gamma_0$ is the parameter, which characterises the nearest-neighbour hopping in a standard
tight-binding approach. According to Haldane \cite{Haldane}
the linearity of bands in graphene (in the vicinity of the $K$ and $K'$ points) implies, that the  dispersion relation (\ref{Eq16q}) is key relation for TLL-behavior of electronic system. Therefore, in the first approximation the electronic system of graphene seemingly has to be considered in the literature to be 2D-Tomonaga-Luttinger liquid system. However, the quite other way is used. It is taken into account, that the charge carriers  behavior in pristine graphene is like to relativistic particles with zero rest mass
and constant velocity $v_F$, equaled to  $\approx 10^6 cms^{-1}$. They are  attributed to massless Dirac
fermions, and  their behaviour is described by the effective
Hamiltonian \cite{Castro}
\begin{equation}\label{Eq17q} 
\hat{H}
= v_F \left[\begin{array} {*{20}c} 0&\hat{p}_x - i\hat{p}_y \\ \hat{p}_x + i\hat{p}_y&0 \end{array}\right] =  v_F \hat{\vec{\sigma}} \hat{\vec{p}},
\end{equation} 
which is equivalent to the Hamiltonian in the Weyl equation for real relativistic particles
with zero rest mass (originally for neutrinos) derived from the Dirac equation.  Therefore, 
the formalism of Tomonaga-Luttinger liquid is not used for description of the physical properties of graphene, despite on  the linear band structure of given system, which from viewpoint of Tomonaga-Luttinger liquid concept is the most appropriate candidate for TLL description.

In favour of foregoing suggestion (on non-TLL description of 2D-systems) indicate also the results of studies of spin-charge separation effect in the spatial dimension $D = 2$ for some specific systems. It has been proposed, in particular,
that the key feature underlying the anomalous behavior
of the cuprate high-$T_c$ materials is precisely a separation
of spin and charge, and concrete scenarios, based on
$Z2$ or $U(1)$ gauge theories, without using of TLL concept, have been put forward \cite{Senthil}.

In the paper \cite{Ardonne} the quantum Hall (qH)
regime, which is relevant for 2D electrons in strong magnetic
fields, is considered. In particular, it is discussed the separation of spin and
charge in given regime. Specifically, it was proposed a series
of paired spin-singlet qH states, of filling fraction
$\nu$ = 2/2m+1. The fundamental
excitations over these states proposed to be spinons (with
spin 1/2 and zero charge) and holons (with zero spin and
fractional charge ${\pm}1/2m+1$, in units of the charge of the
electron). The  statistics of these excitations are
non-abelian, and thereby the paired spin-singlet states
fall in the category of ‘non-abelian qH states’. The prediction of fractional spin-charge separation by spinon-holon mechanism requires, naturally, its experimental confirmation, which, to our knowledge, is  also not provided.
It is interesting, that the more conventional
‘abelian’ spin-singlet qH states like to the Halperin
states with label (m + 1, m + 1, m) do not  predicted even theoretically to be
exhibiting a separation of spin and charge \cite{Ardonne}.

 Further, the key argument for insertion of the notion "Luttinger liquid" itself is in fact  also the simplification, connected (let us  accentuate once again) with linearization of the generic spectrum of particles in neighborhood of Fermi points in k-space. Let us give some details concerning given non-Fermi liquid physics paradigm. Tomonaga's idea \cite{Tomonaga}, 
that the low-energy degrees of freedom of a 1D Fermi gas are completely collective,  has allowed the development of the
"bosonization" technique. At the same time, the conceptual starting point for the bosonization of the Fermi surface is  (let us remember once again) the Luttinger theorem \cite{Luttinger_J}, from here arose the term "Luttinger liquid", introduced by Haldane. However, how it was remarked in \cite{Schmidt}, even for a linear spectrum, the bosonic or fermionic languages may be used equally comfortably and both offer their particular benefits. The advantage of the former is the direct
relation between the bosonic modes and the density response functions. On the other hand, the fermionic
description connects to the well-known physics of the Fermi edge problem. Moreover, they have shown, that
in order to calculate the dynamic response functions in the case of the nonlinearity of the fermion spectrum,
it is convenient to translate the bosonic spin and charge modes into fermionic quasiparticles. So, we see, that if to take into account the only nonlinearity of the generic spectrum of particles in neighborhood of Fermi points in k-space, then the Fermi liquid description of 1D systems becomes to be more convenient, although it remains distorted without including into consideration the electron-phonon interaction. 

When concerning the Landau Fermi liquid theory we accentuate once again, that just oversimplification above indicated has led to divergencies arising in the perturbation theory in 1D-case, that is in LFL theory. However,  it does not means that 1D Fermi liquid description is incorrect in general case.
 We will show, that the concept of adequate description (not distorting the real processes) of 1D correlated electronic systems within the framework of 1D Fermi liquid (FL) can be recovered, at that FL concept can be applied just to 1D carbon NTs, that is to very perspective materials in applications in many branches of modern technology. It is  the aim of the presented work. 

 We will  consider the concept of 1D FL on the example of well known 1D system - \textit{trans}-polyacetylene, that is, it will be in fact the generalization of well known  model, proposed by Su, Schrieffer, Heeger (SSH-model), which, in distinction from LFL and TLL models, takes into account the electron-phonon interaction (however, it does not take into consideration electron-electron correlations in explicit form). The subsequent generalization, for instance, for quasi-1D carbon zigzag shaped nanotubes (CZSNTS) can be  easily  obtained by using of hypercomplex number theory like to description of quantum optics effects, considered in \cite{Dovlatova_Yerchuck}, \cite{Yerchuck_Dovlatova}.

Let us concern briefly the history of spin-charge separation effect.  The idea of spin-charge separation was  explicitly treated  for the first time already in 1974 by Luther and Emery \cite{Luther}
in the context of a continuum limit of the $1D$ electron gas theory.  They have shown, that the Hamiltonian $\hat{H}_{1DEG}$ of the 1D electron gas can
be represented  in the form of
\begin{equation} 
\label{eq1} 
\hat{H}_{1DEG} = \hat{H}_c[\phi_c] + \hat{H}_s[\phi_s] + \hat{H}_{irr}[\phi_c, \phi_s], 
\end{equation} 
where $\hat{H}_c[\phi_c]$ and  $\hat{H}_s[\phi_s]$ are, respectively, the Hamiltonians,
which govern the dynamics of the spin and charge fields,
$\phi_c$ and $\phi_s$, respectively. $\hat{H}_{irr}[\phi_c, \phi_s]$   consists of terms that can be neglected
in the long wave-length limit.  

The related model, which describes spin-charge separation in 1D systems, is the model of the formation of  solitons
with fractional fermion number.  General idea belongs to  Jackiw and
Rebbi \cite{Jackiw}. They have drawn attention to the field theories, especially
in one spatial dimension, which lead solitons' formation
with fractional fermion number. However, the concrete realization of  given idea in condensed matter physics belongs to  Su, Schrieffer, and Heeger \cite{SSH},  \cite{SSH_PRB}.  The  model, proposed by Su, Schrieffer, and Heeger  with spin-charge separation
 to be  the basis phenomenon,   is the model of conjugated organic 1D-conductors.

Specifically, what Su, Schrieffer, and Heeger showed, is that, when an electron
is added to an  neutral \textit{trans}-polyacetylene  chain, it
can break up into two pieces, one of which carries the electron’s charge and the other its spin.  
The real significance of the SSH-soliton model of t-PA consists in  that, that it introduced a new paradigm into
the field, that, in its turn, was  the evidence of triumph of the  model. Triumph of the SSH-model is not occasional. The formulation of the model is very simple from mathematical viewpoint and the simplicity itself is the great advantage of the model. At the same time it demonstrates the  deep physical insight of Su, Schrieffer and Heeger in the field, that was argued in \cite{S_Kivelson}, \cite{Dovlatova_Yerchuck}, \cite{Yerchuck_Dovlatova} and consequence of which is the possibility of the extension of the model. So, in \cite{Yearchuck_PL} was shown, that SSH-model
can be extended even on some polymers with $–C–C–$ ordinary
bonds, which are not possessing by any $\pi$-polyconjugation.

Let us  remark, that  the term, which takes into consideration the static  electron-electron correlations is not presented in SSH Hamiltonian in explicit form, it, in fact, is represented in implicit form. Really, the static electron-electron interaction  can be taken into account in the model  by means of its renormalization into electron-phonon interaction with effective coupling parameter. It was undertaken in  \cite{Zimanyi}, \cite{Voit}. It is very interesting, that the very similar theoretical result on the possibility to renormalize electron-phonon coupling into equivalent electron-electron static interactions was obtained independently  many years later (in 2006) in \cite{Ning}. It was shown, that,  the spin-1/2 Holstein model could be mapped onto the negative-$U$ Hubbard model with an effective dynamical attraction $U_{eff}(\omega)$, dependence of which on the frequency $\omega$ is given by the relation $U_{eff}(\omega) = \lambda/(1 - \omega^2/\omega^2_0)$, where $\lambda$ is the electron-phonon
coupling constant in energetic units, $\omega_0$ is the bare phonon frequency. 

At the same time, although 
the model, used in  \cite{Zimanyi}  is the standard continuum model of a one-dimensional electron gas with short-range (that is screened) electron-electron repulsions, and a nearly
half-filled band, however, the very essential simplification has been done in given model. It consists in linearization of  one electron spectrum about the Fermi surface, that  seems to be oversiplification. Moreover, the idea of full renormalization of electron-electron interaction  into electron-phonon interaction with effective coupling parameter seems to be correct the only partly (see further). It concerns also  in the principle the inverse task \cite{Ning} above cited. The same extended Hubbard model, that in \cite{Zimanyi}, was used in \cite{Voit}, in which the exact bare phonon propagator at the renormalized
electronic energy scale was obtained and used.  It  leads to different physical conclusions concerning the possibility of observation  of charge density wave (CDW) - singlet superconductivity (SS) crossover in t-PA in comparison  with work  \cite{Zimanyi}, where the approximate form  of phonon propagator was
used. Voit has concluded, that
a CDW-SS crossover
does not occur in the interacting SSH model in distinction from the opposite conclusion in \cite{Zimanyi}. Let us remark,  that in both the works above cited  the potential energy of electron-electron correlations is considered to be constant relatively the dimerization coordinate by renormalization procedure, it is also oversimplification (which was mentioned above), since in real physical processes the dimerization coordinate derivative of the potential energy of electron-electron correlations  seems to be the most essential.

The merit of SSH-model, consisting in the choose of the  only dimerization coordinate $u_n$ of the $n$-th $CH$-group, $n = \overline{1,N}$, along chain molecular-symmetry axis $x$ for determination of main physical properties of the material and neglecting by the other five degrees of freedom, that is, the degrees of freedom, which are relevant to the bonds with the directions not coinciding  with chain molecular-symmetry axis direction, seems to be substantial. Given moment was commented in \cite{Dovlatova_Yerchuck} and in \cite{Yerchuck_Dovlatova}. The  possibility to neglect by  five degrees of freedom is the consequence of  general principle, which was proposed by Slater at the earliest stage of quantum physics era already in 1924 \cite{Slater}. It is - "Any
atom may in fact be supposed to communicate with other
atoms all the time it is in stationary state, by means
of virtual field of radiation, originating from oscillators
having the frequencies of possible quantum transitions...". Given idea has  obtained its development in \cite{Dovlatova_Alla_Yerchuck}, where the origin of virtual field of radiation was  clarified. It was shown, that Coulomb field in 1D-systems or 2D-systems
can be quantized, that is, it has the character of radiation field and it can exist without the sources, which
have created given field. Consequently, Slater principle
can be applied to t-PA. In t-PA Coulomb field can be considered to be "virtual" field with propagation direction
the only along t-PA chain. In other words, it produces
preferential direction in atom communication the only in
one direction (to be consequence of quasi-one-dimensionality), and given direction remains to be preferential by
interaction with external EM-field. It explains qualitatively the success of SSH-model in the sense, that degrees
of freedom, realized by bonds, which are not coinciding
with chain molecular-symmetry axis direction, can really
be not taken into consideration for experiments with the
participation of external EM-field and indicating thereby
on deep physical insight of Su, Schrieffer, and Heeger in
the field.

 However, the most merit of SSH-model, which demonstrates a very deep insight of authors in the field, was not commented up to now. In fact, the only given model in the condensed matter physics of dynamic electronic systems takes into consideration in explicit form the generic coupling between operators of creation and annihilation of two quantum fields - between the operators of the field corresponding to electronic subsystem and the operators of the field of lattice deformation system, that is, phonon field. The simplest static analogue of taking into account the  generic coupling between given two fields   is quantum chemistry calculations of the structure of point centers in crystals. It is well known, that by the change of the charge state of any point center in crystal lattice,  the atomic relaxation of neighbourhood host lattice atoms has to be taken into account. In dynamical case it corresponds to phonon absorption or generation. It seems to be evident, that in SSH-model the operators of phonon subsystem are represented through operators of electronic subsystem taking into account given coupling in explicit form. Let us remark, that usually given operators are considered independently on each other, which can lead to distortion of description of real physical processes. 

It seems to be interesting, that there are fundamental qualitative differences by description of spin-charge separation effects in 1D systems between SSH mechanism and  the resulting of TLL
theory Anderson mechanism, applicability of  which to correlated electronic systems seems to be not experimentally confirmed in distinction from SSH mechanism.  The main differerence consists in the role of  phonon effects in spin-charge separation phenomenon.  Let us remember,  that inclusion of electron-phonon interaction in TLL-model suppresses the spinon-holon spin-charge separation effect \cite{Ning} at all. At the same time  electron-phonon interaction plays the essential role for spin-charge separation presence in SSH-model. 

Let us remark, that there is in existing variant of  SSH-model an upper limit on the value of electron-phonon coupling constant. It is consequence of the treatment of electron-phonon coupling to be the linear term in expansion of the only hopping integral of tight-binding model about the undimerized state. Given restriction was discussed in \cite{Rice_M} and the maximum for allowed value of electron-phonon coupling constant $\alpha$  was evaluated to be $\approx 1.27$. We will show, that given restriction can be remitted in developed variant of SSH-model. 

Su, Schrieffer, Heeger \cite{SSH}, \cite{SSH_PRB}   describe mathematically  the chain of t-PA by considering it to be Fermi gas in the sense, that the electron-electron interaction is not taken into consideretion in explicit form, although electron-phonon interaction is taken into account.  We see, therefore, that SSH-model takes, more strictly speaking, the intermediate place between  Fermi gas and Fermi liquid quantum models. 

The main task of our work is  the development of SSH-model within the framework of completely 1D Fermi liquid description, in accordance with   the aim above formulated, and  to clarify in that way, whether the  phenomenon of spin-charge separation, established in SSH-model, will be preserved in  1D quantum Fermi liquid model. 

\section{Results and Discussion}

We will start from Hamiltonian
\begin{equation}
\label{Eq1m}
\hat{\mathcal{H}}(u) = \hat{\mathcal{H}}_{0}(u) + \hat{\mathcal{H}}_{\pi,t}(u) + \hat{\mathcal{H}}_{\pi,u}(u).
\end{equation}
Like to works  \cite{SSH}, \cite{SSH_PRB} we will consider Born-Oppenheimer approximation.
 The first term in 
(\ref{Eq1m}) is  the following
\begin{equation}
\label{Eq2m}
\begin{split}
\hat{\mathcal{H}}_{0}(u) = \sum_{m}\sum_{s}(\frac{\hat{P}_m^2}{2M^*}\hat{a}^+_{m,s} \hat{a}_{m,s} + K u_m^2 \hat{a}^+_{m,s} \hat{a}_{m,s}).
\end{split}
\end{equation}
It represents itself  the sum of  operator of kinetic energy of CH-group motion (the first term in (\ref{Eq2m})) and the  operator of the $\sigma$-bonding energy
(the second term). Coefficient $K$ in (\ref{Eq2m}) is effective $\sigma$-bonds spring constant, $M^*$ is total mass  of CH-group, $u_m$ is configuration coordinate for $m$-th CH-group,  which corresponds to translation of $m$-th CH-group along the symmetry axis $z$ of the chain, $m = \overline{1,N}$, $N$ is number of CH-groups in the chain, $\hat{P}_m$ is operator of impulse, conjugated to configuration coordinate $u_m$, $m = \overline{1,N}$, $\hat{a}^+_{m,s}$, $\hat{a}_{m,s}$ are creation and annihilation operators of creation or annihilation of quasipartile with spin projection $s$ on the $m$-th chain site in  $\sigma$-subsystem of t-PA. 
The second term in 
(\ref{Eq1m}) can be represented in the form of two components and it is

\begin{equation}\begin{split}
\label{Eq3m}
&\hat{\mathcal{H}}_{\pi,t}(u) = \hat{\mathcal{H}}_{\pi,t_0}(u) + \hat{\mathcal{H}}_{\pi,\alpha_1}(u) =\\
&\sum_{m}\sum_{s}[t_0 (\hat{c}^+_{m+1,s} \hat{c}_{m,s} + \hat{c}^+_{m,s} \hat{c}_{m+1,s})  +\\
& (-1)^m 2 \alpha_1 u (\hat{c}^+_{m+1,s} \hat{c}_{m,s} + \hat{c}^+_{m,s} \hat{c}_{m+1,s})],
\end{split}
\end{equation}
where $\hat{c}^+_{m,s}$, $\hat{c}_{m,s}$ are creation and annihilation operators of creation or annihilation of quasiparticle with spin projection $s$ on the $m$-th chain site in  $\pi$-subsystem of t-PA. 
It is the resonance interaction (hopping interaction in tight-binding model approximation) of
 quasiparticles in $\pi$-subsystem of t-PA electronic system, which is considered to be Fermi liquid, and
in which the only constant and linear terms in Taylor series expansion of resonance integral about the dimerized state are  taking into account.

Operator $\hat{\mathcal{H}}(u)$ is invariant under spatial translations with period $2a$, where $a$ is projection of spacing between two adjacent CH-groups in undimerized lattice on chain axis direction, and which is equal to 1.22 $\AA$. It means, that all various wave vectors $\vec{k}$  in $\vec{k}$-space will be in reduced zone with module of $\vec{k}$ in the range $-\frac{\pi}{2a} \leq k \leq \frac{\pi}{2a}$ \cite{SSH_PRB}. It can be considered like to usual semiconductors to be consisting of two subzones - conduction $(c)$ band and valence $(v)$ band. Then it seems to be convenient to represent the operators $\{\hat{c}^+_{m,s}\}$,   $\{\hat{c}_{m,s}\}$, $m = \overline{1,N}$, in the form
\begin{equation}
\begin{split}
\label{Eq5m}
&\{\hat{c}_{m,s}\} = \{\hat{c}^{(c)}_{m,s}\} + \{\hat{c}^{(v)}_{m,s}\},\\
&\{\hat{c}^+_{m,s}\} = \{\hat{c}^{+(c)}_{m,s}\} + \{\hat{c}^{+(v)}_{m,s}\},
\end{split}
\end{equation} related to $\pi-c$- and $\pi-v$-band correspondingly,
and to define $\vec{k}$-space operators
\begin{equation}
\begin{split}
\label{Eq6m}
&\{\hat{c}^{(c)}_{k,s}\} = \{\frac{i}{\sqrt{N}}\sum_{m}\sum_{s}(-1)^{m+1}\exp(-ikma)\hat{c}^{(c)}_{m,s}\},\\
&\{\hat{c}^{(v)}_{k,s}\} = \{\frac{1}{\sqrt{N}}\sum_{m}\sum_{s}\exp(-ikma)\hat{c}^{(v)}_{m,s}\},
\end{split}
\end{equation}
$m = \overline{1,N}$. The principle, like to  MO LCAO is used in fact to build the operators $\{\hat{c}^{(c)}_{k,s}\}$ and $\{\hat{c}^{(v)}_{k,s}\}$, consisting in that the antibonding character of $c$-band orbitals is taken into account by means of factor $i(-1)^{m+1}$.  Inverse to (\ref{Eq6m}) transform is
\begin{equation}
\begin{split}
\label{Eq7m}
&\{\hat{c}^{(c)}_{m,s}\} = \{\frac{1}{\sqrt{N}}\sum_{k}\exp{i[m(ka + \pi) - \frac{\pi}{2}]}\hat{c}^{(c)}_{k,s}\},\\
&\{\hat{c}^{(v)}_{m,s}\} = \{\frac{1}{\sqrt{N}}\sum_{k}\exp(ikma)\hat{c}^{(v)}_{k,s}\},
\end{split}
\end{equation}
$m = \overline{1,N}$. 

The $\sigma$-operators $\{\hat{a}^+_{m,s}\}$ and $\{\hat{a}_{m,s}\}$, $m = \overline{1,N}$ can also be represented in the form like to (\ref{Eq5m}) for $\pi$-operators and analogous to (\ref{Eq6m}),  transforms can be defined. Then the expression for the operator $\hat{\mathcal{H}}_{0}(u)$ can be rewritten
\begin{equation}
\begin{split}
\label{Eq8m}
&\hat{\mathcal{H}}_{0}(u) = \hat{\mathcal{H}}^{\sigma,c}_{0}(u) + \hat{\mathcal{H}}^{\sigma,v}_{0}(u) =
\sum_{m}\sum_{s}(\frac{\hat{P}_m^2}{2M^*} + K u_m^2) \times \\ 
&\frac{1}{N}\sum_{k}(\hat{a}^{+\sigma,c}_{k,s} \hat{a}^{\sigma,c}_{k,s} + \hat{a}^{+\sigma,v}_{k,s} \hat{a}^{\sigma,v}_{k,s}),
\end{split}
\end{equation}
where $\hat{a}^{+\sigma,c}_{k,s}$,  $\hat{a}^{\sigma,c}_{k,s}$ and  $\hat{a}^{+\sigma,v}_{k,s}$, $\hat{a}^{\sigma,v}_{k,s}$ are $\sigma$-operators of creation and annihilation, related to $\sigma$-$c$-band and to $\sigma$-$v$-band correspondingly. The independence of $|u_m|$ on $m$, $m = \overline{1,N}$, means, that the expression  
$(\frac{\hat{P}_m^2}{2M^*} + K u_m^2)$ is independent on $m$. Then we obtain
\begin{equation}
\begin{split}
\label{Eq9m}
\hat{\mathcal{H}}_{0}(u) = \sum_{k}\sum_{s}(\frac{\hat{P}^2}{2M^*} + K u^2)(\hat{n}^{\sigma,c}_{k,s} + 
\hat{n}^{\sigma,v}_{k,s}),
\end{split}
\end{equation}
where
$\hat{n}^{\sigma,c}_{k,s}$ and  
$\hat{n}^{\sigma,v}_{k,s}$ are operators of number of  $\sigma$-quasiparticles in $\sigma$-$c$-band and  $\sigma$-$v$-band correspondingly.

The expression for $\hat{\mathcal{H}}_{\pi,t_0}(u)$ in terms of $\{\hat{c}^{(c)}_{k,s}\}$ and $\{\hat{c}^{(v)}_{k,s}\}$ is coinciding with known corresponding expression in \cite{SSH}, \cite{SSH_PRB} and it is
\begin{equation}
\begin{split}
\label{Eq10m}
\hat{\mathcal{H}}_{\pi,t_0}(u) =  \sum_{k}\sum_{s} 2t_0 \cos ka (\hat{c}^{+(c)}_{k,s}\hat{c}^{(c)}_{k,s} - \hat{c}^{+(v)}_{k,s}\hat{c}^{(v)}_{k,s})
\end{split}
\end{equation}
The expression for the second  part of operator $\hat{\mathcal{H}}_{\pi,t}(u)$ in terms of $\{\hat{c}^{(c)}_{k,s}\}$ and $\{\hat{c}^{(v)}_{k,s}\}$ is also coinciding in its form with known corresponding expression in \cite{SSH}, \cite{SSH_PRB} and it is given by
\begin{equation}
\begin{split}
\label{Eq10ma}
\hat{\mathcal{H}}_{\pi,\alpha_1}(u) =  \sum_{k}\sum_{s} 4 \alpha_1 u \sin ka (\hat{c}^{+(v)}_{k,s}\hat{c}^{(c)}_{k,s} + \hat{c}^{+(c)}_{k,s}\hat{c}^{(v)}_{k,s}), 
\end{split}
\end{equation}
where subscript $\alpha_1$ in Hamiltonian designation indicates on the taking into account the part of electron-phonon interaction, connected with resonance interaction  (hopping) processes.

The expression for the  $\hat{\mathcal{H}}_{\pi,u}(u)$, which describes the part of electron-phonon interaction, determined by interaction  between quasiparticles in Fermi liquid state of $\pi$-subsystem in terms of $\{\hat{c}^{(c)}_{k,s}\}$ and $\{\hat{c}^{(v)}_{k,s}\}$ can be represented in the form 
\begin{equation}
\begin{split}
\label{Eq11m}
\hat{\mathcal{H}}_{\pi,u}(u) = \sum_{k}\sum_{k'}\sum_{s}\alpha_2(k, k',s) \hat{c}^{+(c)}_{k',s}\hat{c}^{+(v)}_{k',s}  \hat{c}^{(v)}_{k,s}\hat{c}^{(c)}_{k,s}.
\end{split}
\end{equation}
The constant independent on $u$ static term, which is determined by electron-electron interaction on different atomic sites in a chain, that is the constant terms in Taylor series expansion of potential energy of electron-electron interaction about the
dimerization coordinate  was omitted in its explicit form from Hamiltonian in given work, in order to establish the role of phonon assisted part. The independent on $u$ static  term is taking, however,  into consideration by calculation of coefficient $\alpha_2(k, k',s)$ in phonon assisted term.

Physically, the identification of linear on displacement $u$ parts of both resonance interaction (hopping) and the pairwise interaction of quasiparticles in $\pi$-subsystem between themselves with electron-phonon interaction is understandable, if to take into account, that by atomic $CH$ group displacements the phonons  are generated, which in its turn can by release of the place on, for instance, $(CH)_m$ group, to deliver the energy and impulse, which are necessary for transfer of the quasiparticle (electron) from adjacent $(m-1)$- or $(m+1)$-position in chain in the case of  resonance interaction (hopping). For the case the pairwise interaction of quasiparticles, it means, that its linear on displacement $u$ part is realized by means of phonon field,  which transfers the energy and impulse from one quasiparticle to another (which can be not inevitable adjacent). Mathematically it can be proved in the following way. The processes of interaction in $c$ ($v$) band can be considered to be independent on each other. It means, that transition probability from the $\langle k_{l,s}|$-state to $\langle k_{j,s}|$-state in $c$-band and from $\langle k'_{l,s}|$-state to  $\langle k'_{j,s}|$-state  in $v$-band, which is proportional to coefficient $\alpha_2(k, k',s)$, can be expressed in the form of product of real parts of corresponding matrix elements, that is in the form
\begin{equation}
\begin{split}
\label{Eq12m}
&\alpha_2(k, k',s) \sim  Re\langle k_{l,s}|\hat{V}^{(c)}|k_{j,s}\rangle Re\langle k'_{l,s}|\hat{V}^{(v)}|k'_{j,s}\rangle = \\
&\sum_{k_{ph}}Re\langle k_{l,s}|\hat{V}^{(c)}|k_{ph}\rangle \langle k_{ph}|k_{j,s}\rangle \times \\
&\sum_{k_{ph}}Re\langle k'_{r,s}|\hat{V}^{(v)}|k_{ph}\rangle\langle k_{ph}|k'_{n,s}\rangle,  
\end{split}
\end{equation} 
where $\hat{V}^{(v)}$ = $V_{0(v)}\hat{e}$ ($\hat{e}$ is unit operator) is the first term in Taylor expansion of pairwise interaction of quasiparticles, for instance, with wave vectors  $k'_{r}$,  $k'_{n}$ and spin projection $s$ in $v$-band, that is, in ground state, $\hat{V}^{(c)}$ = $V_{1(c)} u \hat{e}$ is the second term in Taylor expansion of pairwise interaction in excited state (in c-band), that is, it is   product of configuration coordinate $u$ and coordinate derivative at $u = 0$  of operator of pairwise interaction of quasiparticles with wave vectors $k_{l}$, $k_{j}$ and spin projection $s$   in $c$-band, $k_{ph}$ is phonon wave vector, and the summation is realized over all the phonon spectrum. At that, since the linear density of pairwise interaction is independent on $k$, which is the consequence of translation invariance of the chain, $V_{0(v)}$, $V_{1(c)}$ are constants. Therefore, the  pairwise interaction is considered to be accompanying by process of phonon generation, when electronic quasiparticles are already in excited state, that is,   in $c$-band (retardation effect of phonon subsystem is taken into account).   A number of variants are possible along with process of phonon generation, corresponding to states of electronic quasiparticles in $c$-band above described.  The result will mathematically be quite similar, if to change the energetic place of excitation, that is, if to interchange the role of $c$ and $v$ bands for given process. There seems to be possible the realization of both the stages (that is phonon generation and absorption) for electonic quasiparticles in single $c$ or $v$ band states and simultaneous realization both the stages in both the bands.  For distinctness we will consider  the first variant only. For the simplicity we consider also the processes, in which the spin projection is keeping to be the same. It is evident also,  that in $z$-direction the impulse distribution is quasi-continuous (since the chain has  the macroscopic length $L = N a$).
Consequently, standard way $\sum_{k_{ph}}\rightarrow \frac{L}{2\pi}\int_{k_{ph}}$ can be used. Further, phonon states can be described by wave functions $\langle k_{ph}| = v_0 exp(ik_{ph} z)$, where $z \in [0,L]$, $k_{ph} \in [-\frac{\pi}{2a}$, $\frac{\pi}{2a}$], $v_0$ is constant. Therefore, from (\ref{Eq12m}) we have the expression
\begin{equation}
\begin{split}
\label{Eq13m}
&\alpha_2(k, k',s) = b |v_{0v}|^2 |v_{0c}|^2  V_{0(c)} u V_{0(v)} |\phi_{0cs}|^2 |\phi_{0vs}|^2 \times\\ 
&\frac{N}{2\pi(q_l - q_j)(q_r - q_n)} Re\{\exp[{i(k_l m_l - k_j m_j)a}] \exp{ika}\} \times \\
&Re\{\exp[{i(k'_r m_r - k'_n m_n)a}] \exp {ik'a}\},
\end{split}
\end{equation}
where $|\phi_{0cs}|^2$, $|\phi_{0vs}|^2$ are squares of the modules of the wave functions $|k_{j,s}\rangle$ and $|k'_{j,s}\rangle$ respectively, $k = k_{ph}(q_l - q_j)$, $k' = k'_{ph}(q_r - q_n)$ $q_l, q_j, q_r, q_n \in N$ with additional conditions $(q_l - q_j)a \leq L$, $(q_r - q_n)a  \leq L$, $b$ - is aspect ratio, which in principle can be determined by comparison with experiment. Here
the values $(q_l - q_j)$, $(q_r - q_n)$ determine the steps  in pairwise interaction with phonon participation and they are considered to be fixed. We will consider the processes for which $k = k'$, consequently, $(q_r - q_n)$ =  $(q_l - q_j)$. 

The relation (\ref{Eq13m})
 by $k_{l }m_l = k_{j} m_j$ and by $k_r m_r = k_n m_n$ transforms into the following expression
\begin{equation}
\begin{split}
\label{Eq14m}
&\alpha_2(k, k',s) = b |v_{0v}|^2 |v_{0c}|^2  V_{0(c)} u V_{0(v)} |\phi_{0cs}|^2 |\phi_{0vs}|^2 \times \\
&\frac{N}{2\pi[(q_l - q_j)]^2} \sin ka  \sin k'a.
\end{split}
\end{equation}
Let us designate 
\begin{equation}
\begin{split}
\label{Eq15m}
&b |v_{0v}|^2 |v_{0c}|^2  V_{0(c)}  V_{0(v)} |\phi_{0cs}|^2 |\phi_{0vs}|^2 \times \\
&\frac{N}{2\pi[(q_l - q_j)]^2} = 4 \alpha_2(s)
\end{split}
\end{equation}
Then, taking into account that spin projection $s$ is fixed, the dependence on $s$ can be omitted,  consequently  $\alpha_2(s) =  \alpha_2$. So we have
\begin{equation}
\begin{split}
\label{Eq16m}
&\hat{\mathcal{H}}_{\pi,u}(u) = \\ 
&\sum_{k}\sum_{k'}\sum_{s}4 \alpha_2 u \sin ka  \sin k'a \hat{c}^{+(c)}_{k',s}\hat{c}^{+(v)}_{k',s}  \hat{c}^{(v)}_{k,s}\hat{c}^{(c)}_{k,s}.
\end{split}
\end{equation}

Something otherwise will be treated the participation of the phonons in linear on $u$ part of   pairwise interaction,
if phonon generation is accompanying process of quasiparticle transition from $v$-band into $c$-band, that is when the retardation effect of phonon subsystem can be neglected. It is the case of strong electron-photon interaction, described in \cite{Slepyan_Yerchak}, \cite{Dovlatova_Yerchuck}, \cite{Yerchuck_Dovlatova}. By strong electron-photon interation absorption process of photons is long time process. It is accompanying by quantum Rabi wave packet formation and space propagation, that is by formation of longlived coherent state of joint photon-electron system. In given case
the expression for density of the electron-phonon coupling parameter $\alpha_2(k, k',s)$, which is related to the part of electron-phonon interaction, determined by interaction  between quasiparticles in $\pi$-system Fermi liquid, is the following
\begin{equation}
\begin{split}
\label{Eq17m}
&\alpha_2(k, k',s) \sim  Re\langle k_{l,s}|\hat{V}^{}|k'_{j,s}\rangle = |v_{0v}|^2 |v_{0c}|^2   u V_{1} |\phi_{0s}|^2  \times\\
&\frac{N^2}{[2\pi]^2} 
\int_{k_{ph}}\exp[i(k_{ph} q a - k_{l} m_l a)]  \times\\
&\{\int_{k'_{ph}}\exp[i(k'_{ph} - k_{ph})q'a]  \times\\
&\exp[-i(k'_{ph} q' a - k'_{j} m_n a)]dk'_{ph}\}dk_{ph},
\end{split}
\end{equation}
where $q = m_j - m_l$, $q' = m_r - m_n$ are integers, satisfying foregoing relations, subscrips in left part are omitted, since fixed step is considered.
Then,  taking into account, that in  continuous limit by integration  the modules $k$ and $k'$ of  wave vectors $\vec{k}$ and $\vec{k'}$ are running all the $k$- and $k'$-values in $k$- and $k'$-spaces, we can  designate $(k_{ph} q a - k_{l}m_l a) = ka$, $(k'_{ph} q' a - k'_{j} m_j a) = k'a$ omitting the subscrips. In a result we obtain
\begin{equation}
\begin{split}
\label{Eq18m}
&\alpha_2(k, k',s) \sim  Re\langle k_{l,s}|\hat{V}^{}|k'_{j,s}\rangle = |v_{0v}|^2 |v_{0c}|^2   u V_{1} |\phi_{0s}|^2  \times\\
&\frac{N^2}{[2\pi]^2} (\sin ka  \sin k'a + \cos ka  \cos k'a). \end{split}
\end{equation}
It was taken into account, that by $v$-band $\rightarrow$ $c$-band  transition of  quasiparticle, the impulse of emitted phonon by vibronic system with electronic quasiparticle in $v$-band is equal to the impulse of absorbed phonon by vibronic system with electronic quasiparticle in  $c$-band.

The system of operators $\hat{c}^{+(c)}_{k',s}$, $\hat{c}^{+(v)}_{k',s}$,  $\hat{c}^{(v)}_{k,s}$, $\hat{c}^{(c)}_{k,s}$ corresponds to noninteracting  quasiparticles, and it is understandable, that in the case of  interacting  quasiparticles their linear combination has to be used 
\begin{equation}
\begin{split}
\label{Eq19m}
\left[\begin{array} {*{20}c}  \hat{a}^{(v)}_{k,s} \\  \hat{a}^{(c)}_{k,s} \end{array}\right] = \left[\begin{array} {*{20}c} \alpha_{k,s} & -\beta_{k,s}  \\  \beta_{k,s} & \alpha_{k,s} \end{array}\right] \left[\begin{array} {*{20}c}  \hat{c}^{(v)}_{k,s} \\  \hat{c}^{(c)}_{k,s}  \end{array}\right], 
\end{split}
\end{equation}
where matrix of transformation coefficients $||A||$ is
\begin{equation}
\begin{split}
\label{Eq20m}
||A|| = \left[\begin{array} {*{20}c} \alpha_{k,s} & -\beta_{k,s} \\  \beta_{k,s} & \alpha_{k,s}, \end{array}\right]  
\end{split}
\end{equation}
it is unimodulary matrix with determinant $det\|A\|= \alpha^2_{k,s} + \beta^2_{k,s} = 1$.  Since $det\|A\| \neq 0$, it means, that inverse 
transformation exists and it is given by the matrix
\begin{equation}
\begin{split}
\label{Eq21m}
\|A\|^{-1} = \left[\begin{array} {*{20}c} \alpha_{k,s} & \beta_{k,s}  \\ - \beta_{k,s} & \alpha_{k,s} \end{array}\right].  
\end{split}
\end{equation}
Then we obtain for the Hamiltonian $\hat{\mathcal{H}}_{\pi,\alpha_1}(u)$, which corresponds to SSH one-electron treatment of electron-phonon coupling, the following relation
\begin{equation}
\begin{split}
\label{Eq22m}
&\hat{\mathcal{H}}_{\pi,\alpha_1}(u) = \\
&\sum_{k}\sum_{s}\Delta_k [\alpha^2_{k,s} \hat{a}^{+(v)}_{k,s} \hat{a}^{(c)}_{k,s} - 
\alpha_{k,s} \beta_{k,s} \hat{a}^{+(v)}_{k,s}\hat{a}^{(v)}_{k,s} \\
&+ \beta_{k,s} \alpha_{k,s} \hat{a}^{+(c)}_{k,s}\hat{a}^{(c)}_{k,s} - 
\beta^2_{k,s} \hat{a}^{+(c)}_{k,s} \hat{a}^{(v)}_{k,s} 
+ \alpha^2_{k,s} \hat{a}^{+(c)}_{k,s} \hat{a}^{(v)}_{k,s} \\
&+ \alpha_{k,s} \beta_{k,s} \hat{a}^{+(c)}_{k,s} \hat{a}^{(c)}_{k,s}  -
\beta_{k,s} \alpha_{k,s} \hat{a}^{+(v)}_{k,s}\hat{a}^{(v)}_{k,s} - \beta^2_{k,s} \hat{a}^{+(v)}_{k,s} \hat{a}^{(c)}_{k,s}], 
\end{split}
\end{equation}
where $\Delta_k = 4 \alpha_1 u \sin ka$.

The diagonal part $\hat{\mathcal{H}}^d_{\pi,\alpha_1}(u)$ of operator $\hat{\mathcal{H}}_{\pi,\alpha_1}(u)$  is
\begin{equation}
\begin{split}
\label{Eq23m}
&\hat{\mathcal{H}}^d_{\pi,\alpha_1}(u) = \sum_{k}\sum_{s}2 \Delta_k \alpha_{k,s} \beta_{k,s} (\hat{n}^{(c)}_{k,s} - \hat{n}^{(v)}_{k,s}), 
\end{split}
\end{equation}
where $\hat{n}^{(c)}_{k,s}$ is density of operator of quasiparticles' number in $c$-band,  $\hat{n}^{(v)}_{k,s}$ is density of operator of quasiparticles' number in $v$-band.

 The part of pairwise interaction, which is  linear in displacement coordinate $u$ and leads to participation of the phonons,  is given by 
the Hamiltonian 
\begin{equation}
\begin{split}
\label{Eq24m}
&\hat{\mathcal{H}}_{\pi,u}(u) = \sum_{k}\sum_{k'}\sum_{s} 4 \alpha_2 u \sin ka  \sin k'a \times\\ &(\alpha^2_{k',s} \hat{a}^{+(c)}_{k',s}\hat{a}^{(v)}_{k',s} - \beta^2_{k',s} \hat{a}^{(v)}_{k',s} \hat{a}^{+(c)}_{k',s}\\
&+ \alpha_{k',s} \beta_{k',s} \hat{a}^{(c)}_{k',s} \hat{a}^{+(c)}_{k',s} -
\beta_{k',s} \alpha_{k',s} \hat{a}^{(v)}_{k',s} \hat{a}^{+(v)}_{k',s}) \\
&\times (\alpha^2_{k,s} \hat{a}^{+(c)}_{k,s} \hat{a}^{(v)}_{k,s} - \beta^2_{k,s} \hat{a}^{+(v)}_{k,s} \hat{a}^{(c)}_{k,s}\\
&+ \alpha_{k,s} \beta_{k,s} \hat{a}^{+(c)}_{k,s} \hat{a}^{(c)}_{k,s} - \beta_{k,s} \alpha_{k,s} \hat{a}^{+(v)}_{k,s} \hat{a}^{(v)}_{k,s}).
\end{split}
\end{equation}

The diagonal part $\hat{\mathcal{H}}^d_{\pi,u}(u)$ of operator $\hat{\mathcal{H}}_{\pi,u}(u)$  is
\begin{equation}
\begin{split}
\label{Eq25m}
&\hat{\mathcal{H}}^d_{\pi,u}(u) = 4 \alpha_2 u \sum_{k}\sum_{k'}\sum_{s} \alpha_{k'} \beta_{k'} (\hat{n}^{(v)}_{k',s} - \hat{n}^{(c)}_{k',s}) \\
&\times \alpha_{k,s} \beta_{k,s} (\hat{n}^{(v)}_{k,s} - \hat{n}^{(c)}_{k,s}) \sin k'a \sin ka 
\end{split}
\end{equation}

The Hamiltonian $\hat{\mathcal{H}}_{\pi,t_0}(u)$ in terms of operator variables $\hat{a}^{(c)}_{k,s}$ $\hat{a}^{(v)}_{k,s}$ is

\begin{equation}
\begin{split}
\label{Eq26m}
&\hat{\mathcal{H}}_{\pi,t_0}(u) =  \sum_{k}\sum_{s} 2t_0 \cos ka [(\alpha^2_{k,s}  - \beta^2_{k,s}) (\hat{a}^{+(c)}_{k,s} \hat{a}^{(c)}_{k,s} - \\
&\hat{a}^{+(v)}_{k,s} \hat{a}^{(v)}_{k,s}) - 2 \alpha_{k,s} \beta_{k,s} (\hat{a}^{+(v)}_{k,s} \hat{a}^{(c)}_{k,s} + \hat{a}^{+(c)}_{k,s} \hat{a}^{(v)}_{k,s})]
\end{split}
\end{equation}
The diagonal part $\hat{\mathcal{H}}^d_{\pi,t_0}(u)$ of operator $\hat{\mathcal{H}}_{\pi,t_0}(u)$  is given by the relation

\begin{equation}
\begin{split}
\label{Eq27m}
\hat{\mathcal{H}}^d_{\pi,t_0}(u) = \sum_{k}\sum_{s} \epsilon_k (\alpha^2_{k,s}  - \beta^2_{k,s}) (\hat{n}^{(c)}_{k,s} -
\hat{n}^{(v)}_{k,s}), 
\end{split}
\end{equation}
where $\epsilon_k = 2t_0 \cos ka$. 

The operator transformation for the $\sigma$-subsystem, analogous to (\ref{Eq19m}) shows, that the Hamiltonian $\hat{\mathcal{H}}_{0}(u)$ is invariant under given transformation, that is, it can be represented in the form, given by (\ref{Eq9m}).

To find the values of elements of matrices $\|A\|$ and  $\|A\|^{-1}$, the Hamiltonian $\hat{\mathcal{H}}_{}(u)$
has to be tested for conditional extremum on the variables $\alpha_{k}$, $\beta_{k}$ with condition $\alpha^2_{k,s} + \beta^2_{k,s} = 1$. The corresponding Lagrange function $\hat{\mathfrak{E}}^L_{}(u)$  
is
\begin{equation}
\begin{split}
\label{Eq28m}
&\hat{\mathfrak{E}}^L_{}(u) = \sum_{k}\sum_{s}(\frac{\hat{P}^2}{2M^*} + K u^2)(\hat{n}^{\sigma,c}_{k,s} + 
\hat{n}^{\sigma,v}_{k,s}) \\
&+ \sum_{k}\sum_{s} \epsilon_k (\alpha^2_{k,s}  - \beta^2_{k,s}) (\hat{n}^{(c)}_{k,s} -
\hat{n}^{(v)}_{k,s}) \\
&+ \sum_{k}\sum_{s} 2 \Delta_k \alpha_{k,s} \beta_{k,s} (\hat{n}^{(c)}_{k,s} - \hat{n}^{(v)}_{k,s}) \\
&+ 4 \alpha_2 u \sum_{k}\sum_{k'}\sum_{s} \alpha_{k',s} \beta_{k',s} (\hat{n}^{(c)}_{k',s} - \hat{n}^{(v)}_{k',s}) \alpha_{k,s} \beta_{k,s}\\
&\times(\hat{n}^{(c)}_{k,s} - \hat{n}^{(v)}_{k,s}) \sin k'a \sin ka  + \lambda_{k,s} (\alpha^2_{k,s}  - 1 + \beta^2_{k,s})
\end{split}
\end{equation}
Then, the necessary condition for extremum is determined by  Lagrange equations
\begin{equation}
\begin{split}
\label{Eq29m}
&\frac{\partial{\hat{\mathfrak{E}}^L_{}(u)}}{\partial\alpha_{k}} = 2 \alpha_{k,s}\epsilon_k (\hat{n}^{(c)}_{k,s} - \hat{n}^{(v)}_{k,s}) + 2 \Delta_k  \beta_{k,s} (\hat{n}^{(c)}_{k,s} - \hat{n}^{(v)}_{k,s}) \\
&\times [1 + \frac{\alpha_2}{\alpha_1} \sum_{k'}\sum_{s} \alpha_{k',s} \beta_{k',s} \sin k'a (\hat{n}^{(c)}_{k',s} - \hat{n}^{(v)}_{k',s})] \\
&+ 2 \lambda_{k,s} \alpha_{k,s} = 0,
\end{split}
\end{equation}
\begin{equation}
\begin{split}
\label{Eq30m}
&\frac{\partial{\hat{\mathfrak{E}}^L_{}(u)}}{\partial\beta_{k,s}} = 2 \beta_{k,s}\epsilon_k (\hat{n}^{(v)}_{k,s} - \hat{n}^{(c)}_{k,s}) + 2  \Delta_k  \alpha_{k,s} (\hat{n}^{(c)}_{k,s} - \hat{n}^{(v)}_{k,s}) \\
&\times [1 + \frac{\alpha_2}{\alpha_1} \sum_{k'}\sum_{s} \alpha_{k',s} \beta_{k',s} \sin k'a (\hat{n}^{(c)}_{k',s} - \hat{n}^{(v)}_{k',s})] \\
&+ 2 \lambda_{k,s} \beta_{k,s} = 0
\end{split}
\end{equation}
and
\begin{equation}
\begin{split}
\label{Eq31m}
\frac{\partial{\hat{\mathfrak{E}}^L_{}(u)}}{\partial\lambda_{k,s}} = \alpha^2_{k,s}  - 1 + \beta^2_{k,s} = 0.
\end{split}
\end{equation}
Let us designate
\begin{equation}
\begin{split}
\label{Eq32m}
[1 + \frac{\alpha_2}{\alpha_1} \sum_{k'}\sum_{s} \alpha_{k',s} \beta_{k',s} \sin k'a (\hat{n}^{(c)}_{k',s} - \hat{n}^{(v)}_{k',s})] = \hat{\mathcal{Q}},
\end{split}
\end{equation}
then, passing on to observables in the Lagrange equations (\ref{Eq29m}) - (\ref{Eq31m}), we obtain for $\beta^2_{k,s}$, $\alpha^2_{k,s}$ and for product $\alpha_{k,s} \beta_{k,s}$ the relations
\begin{equation}
\begin{split}
\label{Eq33m}
\beta^2_{k,s} = \frac{1}{2}(1 \pm \frac{\epsilon_k}{\sqrt{\epsilon^2_k + \mathcal{Q}^2 \Delta^2_k}}), 
\end{split}
\end{equation}
\begin{equation}
\begin{split}
\label{Eq34m}
\alpha^2_{k,s} = \frac{1}{2}(1 \mp \frac{\epsilon_k}{\sqrt{\epsilon^2_k + \mathcal{Q}^2 \Delta^2_k}}), 
\end{split}
\end{equation}
\begin{equation}
\begin{split}
\label{Eq35m}
\alpha_{k,s} \beta_{k,s} = \frac{1}{2}\frac{\mathcal{Q} \Delta_k}{\sqrt{\epsilon^2_k + \mathcal{Q}^2  \Delta^2_k}}, 
\end{split}
\end{equation}
where $\mathcal{Q}$ is eigenvalue of operator $\hat{\mathcal{Q}}$.
The equation for factor $\mathcal{Q}$ is
\begin{equation}
\begin{split}
\label{Eq36m}
[1 + \frac{\alpha_2}{2\alpha_1} \sum_{k}\sum_{s}\frac{\mathcal{Q} \Delta_k \sin ka }{\sqrt{\epsilon^2_k + \mathcal{Q}^2 \Delta^2_k}} ({n}^{(c)}_{k,s} - {n}^{(v)}_{k,s})] = \mathcal{Q},
\end{split}
\end{equation}
where superscript {'} is omitted and  ${n}^{(c)}_{k,s}$ is eigenvalue of density  operator of quasiparticles' number in $c$-band,  ${n}^{(v)}_{k,s}$ is eigenvalue of density operator of quasiparticles' number in $v$-band.
It is evident, that at $Q = 1$ in (\ref{Eq33m}) - (\ref{Eq35m}) we have the case of SSH-model. It corresponds to the case, if $\frac{\alpha_2}{\alpha_1} \sum_{k}\sum_{s} \frac{1}{2}\frac{\Delta_k}{\sqrt{\epsilon^2_k +  \Delta^2_k}} \sin ka ({n}^{(c)}_{k,s} - {n}^{(v)}_{k,s})] \rightarrow 0$, which is realized, if $\alpha_2  \rightarrow 0$. Consequently, it seems to be interesting to consider the opposite case, when $|\frac{\alpha_2}{\alpha_1} \sum_{k}\sum_{s} \frac{1}{2}\frac{\Delta_k}{\sqrt{\epsilon^2_k + \Delta^2_k}}\sin ka ({n}^{(c)}_{k,s} - {n}^{(v)}_{k,s})]| \gg 1$. Passing on to continuum limit, in which $\sum_{k}\sum_{s}
\rightarrow 2 \frac{Na}{\pi} \int\limits_0^{\frac{\pi}{2a}}$, and assuming ${n}^{(v)}_{k,s} = 1$, ${n}^{(c)}_{k,s} = 0$, we have integral equation

\begin{equation}
\begin{split}
\label{Eq37m}
\frac{2 N u a \alpha_2}{\alpha_1 \pi t_0} \int\limits_0^{\frac{\pi}{2a}}\frac{\sin^2 ka}{\sqrt{1 - \sin^2 ka [1-(\frac{2 u \mathcal{Q}}{t_0})^2] }} dk = 1.
\end{split}
\end{equation}
In the case $|\frac{2 u \mathcal{Q}}{t_0}| < 1$ the relation (\ref{Eq37m}) can be rewritten in the form

\begin{equation}
\begin{split}
\label{Eq38m} 
&K\left\{\sqrt{1-\left(\frac{2\alpha_1 u \mathcal{Q}}{t_0}\right)^2}  \right\} - E\left\{\sqrt{1-\left(\frac{2\alpha_1 u \mathcal{Q}}{t_0}\right)^2} \right\} = \\
&\frac{\pi [t^2_0 - (2 u \mathcal{Q})^2]}{2 N u \alpha_2},
\end{split}
\end{equation}
where $K\left\{\sqrt{1-(\frac{2\alpha_1 u \mathcal{Q}}{t_0})^2}  \right\}$ and $E\left\{\sqrt{1-(\frac{2 u \mathcal{Q}}{t_0})^2} \right\}$ are complete elliptic integrals of the first and  the second kind, respectively. Expanding given  integrals into the series and restricting by the terms of the second-order of smallness, we obtain

\begin{equation}
\begin{split}
\label{Eq39m} \mathcal{Q} \approx \frac{t_0}{6 u } \sqrt{25 - 32 \frac{t_0 \alpha_1}{N u \alpha_2}}.
\end{split}
\end{equation}
It is evident, that the condition $(\frac{2 u \mathcal{Q}}{t_0}) < 1$  is held true by $\frac{1}{3}\sqrt{25 - 32 \frac{t_0 \alpha_1}{N u \alpha_2}} < 1$.

In the case $|\frac{2 u \mathcal{Q}}{t_0}| > 1$ the relation (\ref{Eq37m}) can be represented in the form
\begin{equation}
\begin{split}
\label{Eq40m}
\int\limits_0^{\frac{\pi}{2}}\frac{\cos^2 y} {\sqrt{1 - \sin^2 y [1-(\frac{t_0}{2 u \mathcal{Q}})^2] }} dy = - \frac{\pi \mathcal{Q} \alpha_1}{\alpha_2 N},
\end{split}
\end{equation}
where $y = \frac{\pi}{2} - ka$. It is equivalent to the equation
\begin{equation}
\begin{split}
\label{Eq41m}
&\left(\frac{t_0}{2 u \mathcal{Q}}\right) F\left\{\frac{\pi}{2},\sqrt{1-\left(\frac{t_0}{2 u \mathcal{Q}}\right)^2}  \right\}\\
&- E\left\{\frac{\pi}{2},\sqrt{1-\left(\frac{t_0}{2 u \mathcal{Q}}\right)^2}  \right\} = \\
&\frac{\pi \mathcal{Q} \alpha_1}{\alpha_2 N}\left[1 - \left(\frac{t_0}{2 u \mathcal{Q}}\right)^2\right],
\end{split}
\end{equation}
where $F\left\{\frac{\pi}{2},\sqrt{1-\left(\frac{t_0}{2 u \mathcal{Q}}\right)^2}  \right\}$ is the complete elliptic integral of the first kind.
The  approximation of elliptic integrals, like to  the approximation, given by  (\ref{Eq39m}), leads to the relation
\begin{equation}
\begin{split}
\label{Eq42m} \mathcal{Q} \approx \frac{-3 \alpha_2 N}{16} \left[ 1 \pm \sqrt{1 + \frac{80 \alpha_1 t_0}{9 N u \alpha_2  }}\right].
\end{split}
\end{equation}

In the case $\frac{2 u \mathcal{Q}}{t_0} = 1$ the relation (\ref{Eq37m}) is 
\begin{equation}
\begin{split}
\label{Eq43m}
\int\limits_0^{\frac{\pi}{2}}\cos^2 y dy = - \frac{\pi \alpha_1 \mathcal{Q}}{\alpha_2 N},
\end{split}
\end{equation}
where $y = \frac{\pi}{2} - ka$.
It is seen, that  in given case the value of parameter $Q$ is calculated exactly and it is
\begin{equation}
\begin{split}
\label{Eq44m}
\mathcal{Q} = \frac{\alpha_2 N}{ 4\alpha_1} 
\end{split}
\end{equation} 
The values of energy of $\pi$-quasiparticles in $c$-band $E_k^{(c)}(u)$ and in  $v$-band $E_k^{(v)}(u)$ can be obtained in the following way
\begin{equation}
\begin{split}
\label{Eq45m}
E_k^{(c)}(u) = \frac{\partial{\mathfrak{E}^L_{}(u)}}{\partial{n^{(c)}_{k,s}}},
E_k^{(v)}(u) = \frac{\partial{\mathfrak{E}^L_{}(u)}}{\partial{n^{(v)}_{k,s}}},
\end{split}
\end{equation}
 where $\mathfrak{E}^L_{}(u)$ is the energy of $\pi$-subsystem of t-PA chain, which is obtained from Lagrange function operator (\ref{Eq28m}) by passing on to observables. Therefore, we have
\begin{equation}
\begin{split}
\label{Eq46m}
&E_k^{(c)}(u) = \epsilon_k (\alpha^2_{k,s}  - \beta^2_{k,s}) + 2 \Delta_k \alpha_{k,s} \beta_{k,s} 
+ 8 \alpha_2 u \sin ka \\
&\times \sum_{k'}\sum_{s} \alpha_{k',s} \beta_{k',s} (\hat{n}^{(c)}_{k',s} - \hat{n}^{(v)}_{k',s})  \sin k'a  \alpha_{k,s} \beta_{k,s} \\
&= \epsilon_k (\alpha^2_{k,s}  - \beta^2_{k,s}) + 2 \Delta_k \alpha_{k,s} \beta_{k,s} \mathcal{Q}
\end{split}
\end{equation}
and 
\begin{equation}
\begin{split}
\label{Eq47m}
&E_k^{(v)}(u) = - \epsilon_k (\alpha^2_{k,s}  - \beta^2_{k,s}) - 2 \Delta_k \alpha_{k,s} \beta_{k,s} 
 - 8 \alpha_2 u \sin ka \\
&\times \sum_{k'}\sum_{s} \alpha_{k',s} \beta_{k',s} (\hat{n}^{(c)}_{k',s} - \hat{n}^{(v)}_{k',s})  \sin k'a  \alpha_{k,s} \beta_{k,s} \\
&=- \epsilon_k (\alpha^2_{k,s}  - \beta^2_{k,s}) - 2 \Delta_k \alpha_{k,s} \beta_{k,s} \mathcal{Q}.
\end{split}
\end{equation}
It is seen from (\ref{Eq46m}) and (\ref{Eq47m}), that $E_k^{(v)}(u) = - E_k^{(c)}(u)$. Taking into account the relations (\ref{Eq33m}) - (\ref{Eq35m}), we obtain
\begin{equation}
\begin{split}
\label{Eq48m}
E_k^{(c)}(u) =  \mp \frac{\epsilon^2_k}{\sqrt{\epsilon^2_k + \mathcal{Q}^2 \Delta^2_k}} + \frac{\mathcal{Q}^2 \Delta^2_k}{\sqrt{\epsilon^2_k + \mathcal{Q}^2  \Delta^2_k}}, 
\end{split}
\end{equation}

\begin{equation}
\begin{split}
\label{Eq49m}
E_k^{(v)}(u) =  \pm \frac{\epsilon^2_k}{\sqrt{\epsilon^2_k + \mathcal{Q}^2 \Delta^2_k}} - \frac{\mathcal{Q}^2 \Delta^2_k}{\sqrt{\epsilon^2_k + \mathcal{Q}^2  \Delta^2_k}}. 
\end{split}
\end{equation}
Therefore, we have two values for the energy of quasiparticles, indicating on the possibility of formation of the quasiparticles of two kinds. Upper sign in the first terms in (\ref{Eq48m}),   (\ref{Eq49m}) corresponds to the quasiparticles with the energy
\begin{equation}
\begin{split}
\label{Eq50m}
&E_k^{(c)}(u) =   \frac{\mathcal{Q}^2 \Delta^2_k  - \epsilon^2_k}{\sqrt{\epsilon^2_k + \mathcal{Q}^2  \Delta^2_k}},\\ 
&E_k^{(v)}(u) =   \frac{\epsilon^2_k - \mathcal{Q}^2 \Delta^2_k}{\sqrt{\epsilon^2_k + \mathcal{Q}^2  \Delta^2_k}}
\end{split}
\end{equation}
in $c$-band and $v$-band respectively. Lower sign in the first terms in (\ref{Eq48m}),   (\ref{Eq49m}) corresponds to  the quasiparticles with the energy
\begin{equation}
\begin{split}
\label{Eq51m}
&E_k^{(c)}(u) =   \sqrt{\epsilon^2_k + \mathcal{Q}^2  \Delta^2_k},\\ 
&E_k^{(v)}(u) =   - \sqrt{\epsilon^2_k + \mathcal{Q}^2  \Delta^2_k}
\end{split}
\end{equation}
in $c$-band and $v$-band respectively.
The quasiparticles of the second kind  at  $\mathcal{Q} = 1$ are the same quasiparticles, that were obtained in \cite {SSH_PRB}. 

We have used the only necessary condition for extremum of the functions $E(\alpha_{k,s}, \beta_{k,s}, \lambda_{k,s})$. It was shown in \cite{Yerchuck_Dovlatova}, that for the  SSH-model  
the  sufficient conditions for the minimum are substantial, they change the role of both solutions.  Sufficient conditions for the minimum of the functions $E(E(\alpha_{k,s}, \beta_{k,s}, \lambda_{k,s})$
 can be  obtained by standard way, which was used in \cite{Yerchuck_Dovlatova}. It consist in that, that the second differential of  the energy being to be the function of  three variables  ${\alpha}_{k,s}$,  ${\beta}_{k,s}$ and  
 $\lambda_{k,s}$  has to be positively defined quadratic form. From the condition of positiveness of three principal minors of quadratic form coefficients we obtain  the following three sufficient conditions for the energy minimum

\paragraph{The first condition}

The first condition is
\begin{equation}
\label{Eq52m}
\begin{split}
&\{ \epsilon_k (1 - \frac{\epsilon_k}{\sqrt{\epsilon^2_k + \mathcal{Q}^2  \Delta^2_k}}) < \frac{(\mathcal{Q}\Delta_k)^2}{\sqrt{\epsilon^2_k + \mathcal{Q}^2  \Delta^2_k}} | 
({n}^c_{k,s} - {n}^v_{k,s}) < 0\}, \\
&\{\epsilon_k (1 - \frac{\epsilon_k}{\sqrt{\epsilon^2_k + \mathcal{Q}^2  \Delta^2_k}}) > \frac{(\mathcal{Q}\Delta_k)^2}{\sqrt{\epsilon^2_k + \mathcal{Q}^2  \Delta^2_k}} | ({n}^c_{k,s} - {n}^v_{k,s}) > 0 \}
\end{split}
\end{equation}
for the solution which coincides with SSH-solution at the value $\mathcal{Q} = 1$ (SSH-like solution) and
\begin{equation}
\label{Eq53m}
\begin{split}
&\{\epsilon_k (1 + \frac{\epsilon_k}{\sqrt{\epsilon^2_k + \mathcal{Q}^2  \Delta^2_k}}) < \frac{(\mathcal{Q}\Delta_k)^2}{\sqrt{\epsilon^2_k + \mathcal{Q}^2  \Delta^2_k}} | ({n}^c_{ks} - {n}^v_{ks}) < 0 \},\\
&\{\epsilon_k (1 + \frac{\epsilon_k}{\sqrt{\epsilon^2_k + \mathcal{Q}^2  \Delta^2_k}}) > \frac{(\mathcal{Q}\Delta_k)^2}{\sqrt{\epsilon^2_k + \mathcal{Q}^2  \Delta^2_k}} | ({n}^c_{ks} - {n}^v_{ks}) > 0 \}
\end{split}
\end{equation}
 for the additional solution. It is seen, that the first condition is realizable for the quasiparticles of both the kinds, at that for both near equilibrium  state $({n}^c_{ks} - {n}^v_{ks} < 0)$  and  strongly nonequilibrium state $(n^c_{ks} - {n}^v_{ks} > 0$.  

\paragraph{The second condition}

The second condition is the same for both the solutions and it is
\begin{equation}
\label{Eq54m}
 (\frac{\epsilon^2_k}{\sqrt{\epsilon^2_k + \mathcal{Q}^2  \Delta^2_k}} - 2\frac{(\mathcal{Q}\Delta_k)^2}{\sqrt{\epsilon^2_k + \mathcal{Q}^2  \Delta^2_k}})^2 - \epsilon^2_k   + \frac{1}{4} (\mathcal{Q}\Delta_k)^2 > 0 
\end{equation}
It  is realizable for the quasiparticles of both the kinds.

\paragraph{The third condition}

For the SSH-like solution we have
\begin{equation}\label{Eq55m}
(3\frac{(\mathcal{Q}\Delta_k)^2}{\sqrt{\epsilon^2_k + \mathcal{Q}^2  \Delta^2_k}} + 4\frac{\epsilon^2_k}{\sqrt{\epsilon^2_k + \mathcal{Q}^2  \Delta^2_k}})({n}^c_{ks} - {n}^v_{ks}) > 0. 
\end{equation}
It  means,  that   SSH-like solution is unapplicable for the description of standard processes, passing near equilibrium state by any parameters. The quasiparticles, described by   SSH-like solution, can be created the only in strongly nonequilibrium state with inverse                                                                                                   
population of the levels in $c$- and $v$-bands. At the same time  the solution, which corresponds to upper signs in (\ref{Eq48m}), has to satisfy  to the following condition
\begin{equation}
\label{Eq56m}
(3\frac{(\mathcal{Q}\Delta_k)^2}{\sqrt{\epsilon^2_k + \mathcal{Q}^2  \Delta^2_k}} - 4\frac{\epsilon^2_k}{\sqrt{\epsilon^2_k + \mathcal{Q}^2  \Delta^2_k}})({n}^c_{ks} - {n}^v_{ks}) > 0, 
\end{equation}
which can be realized  both in near equilibrium and in strongly nonequilibrium states of the $\pi$-subsystem of t-PA, which is considered to be quantum Fermi liquid.

\paragraph{Ground State of  $t$-PA chain}

The continuum limit for the ground state energy of the $t$-PA chain with SSH-like quasiparticles will coincide with known solution \cite{SSH_PRB}, if to  replace $\Delta_k \mathcal{Q} \rightarrow \Delta_k$.  Let us calculate  the ground state energy $E^{[u]}_0(u)$ of the $t$-PA chain with  quasiparticles' branch, which is stable near equilibrium. Taking into account, that in ground state ${n}^c_{k,s} = 0$, ${n}^v_{k,s} = 1$, in the continuum limit we have
\begin{equation}
\label{Eq57m}
E^{[u]}_0(u) = - \frac{2N a}{\pi}\int\limits_0^{\frac{\pi}{2a}} \frac{(\mathcal{Q}\Delta_k)^2 - 
\epsilon^2_k}{\sqrt{(\mathcal{Q}\Delta_k)^2 + 
\epsilon^2_k}}dk + 2NKu^2,
\end{equation}
then, calculating the integral and using the complete elliptic integral of the first kind $F(\frac{\pi}{2}, 1 - z^2)$ and the complete elliptic integral of the second kind
$E(\frac{\pi}{2}, 1 - z^2)$  we obtain
\begin{equation}
\label{Eq58m}
\begin{split}
&E^{[u]}_0(u) =  \frac{4Nt_0}{\pi}\{F(\frac{\pi}{2}, 1 - z^2) + \\
&\frac{1 + z^2}{1 - z^2}[E(\frac{\pi}{2}, 1 - z^2) - F(\frac{\pi}{2}, 1 - z^2)]\} + 2NKu^2, 
\end{split}
\end{equation}
where
$z^2 = \frac{2\mathcal{Q}\alpha_1 u}{t_0}$.
Approximation of ({\ref{Eq58m}}) at $z \ll 1$ gives
\begin{equation}
\label{Eq59m}
\begin{split}
&E^{[u]}_0(u) = N \{\frac{4t_0}{\pi} - \frac{6}{\pi}\ln\frac{2t_0}{\mathcal{Q}\alpha_1 u} \frac{4 (\mathcal{Q}\alpha_1)^2 u^2}{t_0} + \\
&\frac{28 (\mathcal{Q}\alpha_1)^2 u^2}{\pi t_0} + ...\} + 2NKu^2.
\end{split}
\end{equation}
It is seen from (\ref{Eq59m}), that the energy of quasiparticles, described by   solution, which corresponds to upper signs in (\ref{Eq48m}) has the form of Coleman-Weinberg potential with two minima at the values of dimerization coordinate $u_0$ and $-u_0$ like to the energy of quasiparticles, described by   SSH-solution \cite{SSH_PRB}. It is understandable, that subsequent consideration, including electrically neutral S=1/2 soliton formation and electrically charged spinless soliton formation, that is the appearence of the phenomenon of spin-charge separation,  by Fermi liquid description of 1D systems  will be coinciding in its  mathematical form with starting  SSH-model. 

\section{Conclusions}

The possibility to describe the physical properties of  (quasi)-1D systems  within the frames of (quasi)-1D quantum Fermi liquid including the mechanism of appearence of the most prominent feature of (quasi)-1D systems - the phenomenon of spin-charge separation - is proved. Thus, the consideration of (quasi)-1D systems within quantum Fermi liquid concept is recovered.

It is shown, that all qualitative conclusions of the model proposed in \cite{SSH_PRB} are holding in quantum Fermi-liquid consideration of $\pi$-electronic subsystem of t-PA chain (instead of formal Fermi-gas consideration) for the quasiparticles, corresponding to the second-branch-solution. It seems to be substantial, that Fermi-liquid treatment of electron-phonon interaction extends the applicability limits of SSH-model of  (quasi)-1D  conjugated conductors, allowing its use  in the case of strong electron-phonon interaction. 

It is shown, that the mechanism of the phenomenon
of spin-charge separation in of (quasi)-1D Fermi-liquid is topological soliton mechanism, being to be analogous to mechanism proposed by Jackiw and Rebbi. It means, like to SSH-model, that when an electron
is added to an  neutral \textit{trans}-polyacetylene chain (or similar (quasi)-1D system), it
can break up into two pieces, one of which carries the
electron’s charge and the other its spin. Given result
bears a clear family relation with the phenomenon
of  spin-charge separation in the 1D electron gas theory of Luther and Emery \cite{Luther},
but it is quite different from Anderson spinon-holon mechanism.

 The results obtained allow  to make more accurate and to correct the prevalent viewpoint, that spin-charge separation effect is indication on non-Fermi-liquid behavior of electronic systems and that it can be  reasonably described within the framework of Tomonaga-Luttinger liquid theory only. Given 
viewpoint is true the only for the systems with  the function of the  electronic energy in $k$-space, which is characterised by the absence of extremum in  the dependence on $k$. Generally from mathematical viewpoint, Tomonaga-Luttinger  liquid behaviour can be observed independently on the dimensionality for the systems, for which the energy at
Fermi surface is not extremal and, consequently, the linear term has to be
preserved in its Taylor expansion about the Fermi surface points. At the same time, it is argued, that the model of Tomonaga-Luttinger liquid in its existing form  seems to be not sufficiently correct mapping of real processes, since it does not take into account the electron-phonon interaction, which always takes place by the change of charge state   in any 1D system. Moreover, inclusion of the electron-phonon interaction term in Hamiltonian leads to strong decrease of the display of main peculiarity of TLL - spinon-holon spin charge separation - upto its full absence. Our analysis of experimental works shows, that the presence of spinon-holon spin charge separation effect remains to be experimentally unproved.

The properties (including the possibility of the observation of the phenomenon
of spin-charge separation) of many physical (quasi)-1D systems can be described within the framework of  (quasi)-1D quantum Fermi-liquid proposed. For instance, the model proposed seems to be working model for rather  wide
class of conjugated organic conductors.
 
The model of (quasi)-1D Fermi liquid   allows to extend the limits of the applicability of SSH-model for description of the 1D-systems with both strong electron-phonon interaction and (or) strong electron-photon interaction by arbitrary electron - electron interaction. 

It seems to be especially significant, that (quasi)-1D quantum Fermi liquid model proposed can be easily generalised for description of the properties of quasi-1D carbon nanotubes, which are perspective materials for nanoelectronics, spintronics and for the number of the other practical applications.

Practical significance of the model proposed consists also in the clarification of the nature of charge and spin carriers and in the clarification of  the origin of mechanisms of quasiparticles' interaction in  the (quasi)-1D-materials, that is, it can be theoretical base for elaboration of the devices  of nanoelectronics, spintronics  and objects of the other nanotechnology branches.

\end{document}